# Spatial prediction of apartment rent using regression-based and machine learning-based approaches with a large dataset

Takahiro Yoshida[1*] and Hajime Seya[2]

[1] Graduate School of Engineering, The University of Tokyo, Japan
[2] Graduate School of Engineering, Kobe University, Japan
* Corresponding author. Address: 7-3-1 Hongo, Bunkyo-ku, Tokyo 113-8656, Japan.
E-mail: yoshida.takahiro@up.t.u-tokyo.ac.jp

*Abstract*

Employing a large dataset (at most, the order of $n = 10^6$), this study attempts enhance the literature on the comparison between regression and machine learning (ML)-based rent price prediction models by adding new empirical evidence and considering the spatial dependence of the observations. The regression-based approach incorporates the nearest neighbor Gaussian processes (NNGP) model, enabling the application of kriging to large datasets. In contrast, the ML-based approach utilizes typical models: extreme gradient boosting (XGBoost), random forest (RF), and deep neural network (DNN). The out-of-sample prediction accuracy of these models was compared using Japanese apartment rent data, with a varying order of sample sizes (i.e., $n = 10^4, 10^5, 10^6$). The results showed that, as the sample size increased, XGBoost and RF outperformed NNGP with higher out-of-sample prediction accuracy. XGBoost achieved the highest prediction accuracy for all sample sizes and error measures in both logarithmic and real scales and for all price bands (when $n = 10^5$ and $10^6$). A comparison of several methods to account for the spatial dependence in RF showed that simply adding spatial coordinates to the explanatory variables may be sufficient.

*Acknowledgment*
This work was supported by JSPS KAKENHI Grant Numbers: 21K13153, 21H01447, and 18H03628. We used "LIFULL HOME'S Dataset" provided by LIFULL Co., Ltd. via IDR Dataset Service of National Institute of Informatics (https://doi.org/10.32130/idr.6.0). We thank Daisuke Murakami of the Institute of Statistical Mathematics, Hayato Nishi of the University of Tokyo, and attendees of The XV World Conference of Spatial Econometrics Association (May 26–28, 2021; Online) for providing useful comments.

# 1. Introduction

Online automatic real estate price estimation services, such as Zestimate[1] (a service of the Zillow Group in the United States), are increasing in popularity. Seya and Shiroi (2021) reported that accurate price assessments and predictions are crucial for real estate agents, as well as end users. Considering the perspective of the agency, reducing appraisal costs and improving transparency are advantageous. Meanwhile, the perspective of end users involves the improvement of information asymmetry between real estate agencies and end users to a certain extent. Massive property data and statistics- or machine learning (ML)-based real estate sales and rent price prediction methods are the means of support for agents and users.

Traditionally, regression approaches are employed to prices of real estate and rent, although automated assessment of real estate sales and rent prices using techniques involving massive data and ML-based has garnered attention (Abidoye and Chan, 2017; Čeh et al., 2018). Efron (2020) reported regression-based approaches being typically used for prediction as well as *attribution*, that is, the individual predictors being assigned significance (i.e., significance testing). However, combining weak learners in ML-based approaches (e.g., random forest (RF) or extreme gradient boosting [XGBoost]) is not effective in the case of attribution. Thus, regression-based approaches can offer advantages. However, considering pure prediction, simple functional forms, such as linear, logarithmic, and Box-Cox, commonly employed in regression-based approaches, may be inadequate for capturing the nonlinearity of the data[2]. Therefore, examining the extent of the difference in prediction accuracy from ML-based methods is crucial.

Constructing prediction models of real estate sales or rent prices results in certain challenges in accommodating factors such as neighborhood quality as explanatory variables (covariates) (Dubin, 1988). Hence, considering the spatial dependence inherent in the data is important (Pace and LeSage, 2004; Hayunga and Pace, 2010). In geo-(spatial) statistics, regression-based kriging was established to incorporate spatial dependence among error terms, typically applying a Gaussian process (GP) to the errors (Cressie and Wikle, 2011). Certain studies, such as James et al. (2005), Bourassa et al. (2010), and Seya et al. (2011), reported that high predictive accuracy was offered by the kriging approach compared to multiple regression models (ordinary least squares [OLS]) in the property-related literature. Because the OLS model structure is straightforward, it enables parameter determination using relatively small samples. However, with kriging, as the price information of neighboring properties is reflected in the predicted results through spatial dependence, it results in a situation different from that of OLS. That is, several prediction benefits of increasing the sample size exist (Seya and Shiroi, 2021). In contrast, in the case of the ML approach, research that attempts to introduce spatial dependency into the model remains in its early stages, although some interesting studies have recently been conducted (e.g., Saha et al., 2020; Iranzad et al., 2021).

---

[1] https://www.zillow.com/
[2] Certainly, it is possible to use more flexible semiparametric or nonparametric functional forms (Seya et al., 2011).

This study aims to supplement the literature with the addition of new empirical evidence via comparison of regression- and ML-based rent price prediction models used on a large dataset (at most, in the order of $n = 10^6$), and considering the spatial dependence among observations as an expansion of that reported by Seya and Shiroi (2021). The former approach required kriging. However, OLS was also employed to set a general benchmark. Moreover, with increasing sample size (e.g., when $n = 10^5$), application of kriging directly becomes difficult, requiring $O(n^3)$ computational cost to invert the variance–covariance matrix. Therefore, we considered the nearest neighbor Gaussian processes (NNGP) model, allowing the application of kriging to massive data via sparse approximation (Datta et al., 2016; Finley et al., 2017; Zhang et al., 2019). Although exist several methods exist to conduct spatial statistical modeling with big data (Yamagata and Seya, 2019; Banerjee, 2020), NNGP is reliable owing to it consistent competitive results in comparative studies (Heaton et al., 2019). Furthermore, for the latter, certain representative models, namely, RF, XGBoost, and deep neural network (DNN) were employed.

Several trials have compared and investigated the predictive accuracy of real estate sales and rent prices between regression- and ML-based approaches. However, certain limitations should be overcome, including [1] small sample sizes, [2] disregard for spatial dependence, and [3] tailored and ad hoc hyperparameter settings. Hence, in this study, we [1] examined different and relatively large sample sizes ($n = 10^4$, $10^5$, and $10^6$), [2] considered spatial dependence, and [3] finely calibrated the hyperparameters via cross-validation.

Further, this study employed the LIFULL HOME's dataset[3] to evaluate monthly residential apartment rental prices in Japan[4] for empirical evidence. This dataset was also used by Seya and Shiroi (2021) and comprises rental property cross-sectional data and image data up to September 2015. The rental property cross-sectional data include rent, lot size, location (i.e., municipality, zip code, nearest station, and time consumed to walk to the nearest station), year it was built, layout of the room, building structure, and equipment for 5.33 million properties throughout Japan. Meanwhile, the image data comprise 83 million pictures that outline the floor plans and details about the interiors for every property. This study employed only the former data.

Among the 5.33 million properties, 4,588,632 properties were retained after the missing data was excluded. Thereafter, $n = 10^4$, $10^5$, and $10^6$ properties were randomly sampled from the cleaned data. Subsequently, the regression-based (OLS and NNGP) and ML-based approaches (RF, XGBoost, and DNN) were compared via a validation process while considering the difference in sample size and out-of-sample predictive accuracy of rent prices.

The remainder of this paper proceeds as follows. Section 2 presents a short review of existing literature. Section 3 explains the models used in this comparison study. Further, Section 4 details the results obtained from the comparative analysis using the LIFULL HOME dataset. Finally, the concluding remarks, along with the scope for future research, are presented in Section 5.

## 2. Literature review

This section presents a review of the literature regarding the prediction of real estate sales and rent prices. Studies have postulated that spatial regression models exhibit a high predictive accuracy compared to the OLS model (e.g., James et al., 2005). Seya et al. (2011) examined the performance of various spatial prediction models that considered spatial dependence by employing a dataset comprising apartment rents of 23 wards in Tokyo for empirical comparison. They showed the benefit of considering spatial dependence in the error term (e.g., kriging, geoadditive model, and spatial error model) or regression coefficients (e.g., geographically weighted regression (GWR) model). However, a limitation was the small size (i.e., 529 for parameter estimation and 150 for validation).

---

[3] https://doi.org/10.32130/idr.6.0
[4] LIFULL Co., Ltd. provided this to the researchers free of charge through the National Institute of Informatics.

Geostatistical models (kriging) and spatial econometric models are extensively used to consider spatial dependence among errors. Many studies have applied both methods to hedonic price modeling. However, for the purpose of spatial (i.e., out-of-sample) prediction, the former, which requires no spatial weight matrix, is more natural and flexible (Tsutsumi and Seya, 2009). However, a comparison by Seya et al. (2011) revealed that the differences in the predictive accuracy between the kriging and spatial econometric models are negligible, compared to the differences between OLS and kriging. For kriging, application to massive data, on the order of a million, can be achieved by considering various approximations (Heaton et al., 2019).[5]

Various methods have been developed to model spatial dependence among regression coefficients in different fields, including geography, statistics, and ML (e.g., Brunsdon et al., 1998; Gelfand et al., 2003; Murakami et al., 2017; Dambon et al., 2020). Because the housing market is often segmented, the local model (i.e., spatially varying coefficient (SVC) model) can be applied. Hence, SVC models have been employed for hedonic price modeling in several studies. However, their application to massive amounts of data remains in its nascent stages (Li and Fotheringham, 2020; Murakami et al., 2020). For instance, the scalable GWR model, proposed by Murakami et al. (2020), was applied to our dataset. However, the parameter estimation procedure (i.e., the bandwidth selection procedure) was not completed within 24 h when $n = 10^5$. Thus, application of the model to a dataset with a sample size of $n = 10^6$ and above is difficult.[6]

Several attempts have been made to achieve the results based on similar motivation. Seya and Shiroi (2021) reviewed studies that employed neural network (NN) methods. Valier (2020) reported that 57 cases are available wherein ML-based models, including NN, were more accurate in predicting the values than the 13 cases wherein regression performed better. Zurada et al. (2011) suggested that, although many recent studies have compared regression with artificial intelligence (AI)-based methods in the context of mass appraisal, useful comparison of the published results is a challenge because the models in many studies were built by considering relatively small samples. Therefore, for a more comprehensive comparative study, a dataset containing over 16,000 sales transactions was used. They found that non-traditional regression-based methods performed better in all simulation scenarios, specifically with homogeneous datasets. However, AI-based methods performed well with less homogeneous datasets under certain simulation scenarios. Seya and Shiroi (2021), upon which the present study was built, compared the performances of OLS, NNGP, and DNN. They found that, with an increase in sample size from $n = 10^4$ to $10^6$, the DNN's out-of-sample predictive accuracy approaches that of NNGP and is nearly equal in the order of $n = 10^6$. However, in terms of both higher- and lower-end predictive accuracy for which rent prices deviate from the median, DNN may have better results than NNGP. Seya and Shiroi (2021) have a clear limitation in that they only used the DNN method to represent the ML approach.

---

[5] See Yamagata and Seya (2019) for the application of spatial econometric models to big data.
[6] In addition, for $n = 10^4$, the prediction accuracy was no better than that of NNGP (mean absolute error [MAE]: 0.1400).

Several studies have employed tree-based techniques to realize the ML approach. Pace and Hayunga (2020) examined tree-based techniques, including classification and regression trees (CART) (Breiman et al., 1984), boosting (Schapire, 1990), and bagging (Breiman, 1996). Further, they compared these techniques to the spatio-temporal linear model (Pace et al., 1998), considering over 80,000 real estate prices in the United States. Bagging was found to work well and could yield lower out-of-sample residuals than global spatiotemporal methods; however, its performance was poorer than local spatiotemporal methods. Mayer et al. (2019) used a large dataset consisting of over 123,000 single-family houses sold in Switzerland between 2005 and 2017. They reported that the gradient boosting (GB) approach performed far better than the other methods. It was followed by mixed effects regression, the NN method, and the RF approach in terms of performance. Based on the online housing platform, Ming et al. (2020) used 33,224 pieces of data reflecting Chengdu housing rentals in China. They empirically compared the predictive performance of RF, LightGBM, and XGBoost and found that XGBoost performed the best. Ho et al. (2021) used three ML algorithms, namely, support vector machine (SVM), RF, and GB, to appraise property prices. They applied these methods to examine a data sample of approximately 40,000 housing transactions over a period of over 18 years in Hong

Kong and then compared the results of these algorithms. They found that RF and GBM outperformed the SVM in terms of predictive power.

## 3. Models

This section introduces regression- and ML-based approaches to the spatial prediction employed in this study.

### 3.1. Regression-based approaches

#### 3.1.1. NNGP

Consider $D$ as the spatial domain and $s$ the coordinate position (X, Y). The spatial regression model, also referred to as the spatial process model, can be expressed as (Banerjee et al., 2014; Yamagata and Seya, 2019)

$$y(s) = m(s) + w(s) + \varepsilon(s), \quad \varepsilon(s) \sim N(0, \tau^2), \tag{1}$$

where $y(s)$ is the spatial process for real estate rental prices, which is decomposed to $m(s), w(s)$, and $\varepsilon(s)$. Further, $\tau^2$ represents a variance parameter termed as a nugget representing the micro-scale variation and measurement error (Cressie, 1993). Typically, it is assumed that $m(s) = x(s)'\beta$, where $x$ is an explanatory variable vector at point $s$, and $\beta$ is the corresponding regression coefficient vector. Further, $w(s)$ is assumed to follow the GP: $w(s) \sim GP(0, C(\cdot, \cdot | \theta))$, with the mean being zero and the covariance function being $C(\cdot, \cdot | \theta)$ (where $\theta$ is a parameter vector that typically includes the parameter $\phi$ [where $1/\phi$ is called the range]; it controls the range of the influence of spatial dependence, and the parameter $\sigma^2$ represents the variance of the spatial process and referred to as the partial sill). Finally, $\varepsilon(s)$ is an uncorrelated pure error term.

For sample obtained at points $s_1, \ldots, s_n$, with $y(s_i)$ and $x(s_i)$ denoting the dependent variable and explanatory variables observed at location $s_i$, we obtain $w = (w(s_1), w(s_2), \ldots, w(s_n))'$ from the multivariate Gaussian distribution: $w \sim N(0, C(\theta))$. Here, $0$ is an $n \times 1$ vector of zeros, and the element of the $n \times n$ matrix $C(\theta)$ is expressed as $C(s_i, s_j | \theta)$ $(i = 1, \ldots, n; j = 1, \ldots, n)$. Further, $y \sim N(X\beta, \Lambda(\tau^2, \theta))$ can be expressed as the spatial process model, where $\Lambda(\tau^2, \theta) = C(\theta) + \tau^2 I$, where $I$ is an $n{\times}n$ identity matrix.

The prediction of the response $y(\boldsymbol{s}_0)$ at a particular point $\boldsymbol{s}_0$ is called kriging.[7] For the kriging predictor, the inverse of the $n{\times}n$ variance–covariance matrix $\boldsymbol{\Lambda}$ is required. Thus, a cost of $O(n^3)$ is accrued for the computation, which on the order of $n = 10^5$ with a standard personal computer environment poses a challenge. Hence, various approaches are available for approximating the spatial process $\boldsymbol{w}(\boldsymbol{s})$ (e.g., Heaton et al., 2019; Banerjee, 2020). This study employs the NNGP model (Datta et al., 2016), which was originally proposed by Vecchia (1988). The joint density of the spatial process $\boldsymbol{w}$ (the full GP) is expressed as the product of conditional densities, that is, $p(\boldsymbol{w}) = p(w(\boldsymbol{s}_1))\prod_{i=2}^{n} p(w(\boldsymbol{s}_i)|w(\boldsymbol{s}_1), \ldots, w(\boldsymbol{s}_{i-1}))$.[8] Thereafter, Datta et al. (2016) assumed the following approximation for this joint density:

$$\tilde{p}(\boldsymbol{w}) = p(w(\boldsymbol{s}_1))\prod_{i=2}^{n} p\left(w(\boldsymbol{s}_i)\big|w(N(\boldsymbol{s}_i))\right), \qquad (2)$$

where $N(\boldsymbol{s}_i)$ is a neighbor set of $\boldsymbol{s}_i$ and serving as the $k$-nearest neighbors of $\boldsymbol{s}_i$ in NNGP. Thus, the complete GP is approximated by the NNGP and can be expressed as a joint density using the nearest neighbors. Further, Datta et al. (2016) demonstrated that the approximation of Eq. (2) leads to an approximation of the precision matrix $\boldsymbol{C}^{-1}$ to $\widetilde{\boldsymbol{C}}^{-1}$, as expressed as follows:

$$\widetilde{\boldsymbol{C}}^{-1} = \left(\boldsymbol{I} - \widetilde{\boldsymbol{A}}\right)'\boldsymbol{D}^{-1}(\boldsymbol{I} - \widetilde{\boldsymbol{A}}), \qquad (3)$$

where $\widetilde{\boldsymbol{A}}$ is a sparse and strictly lower triangular matrix, with its diagonal elements represented by zeros, with non-zero entries at most $k$-entries in each row. Further, $\boldsymbol{D} = \mathrm{diag}(d_{ii})$ is a diagonal matrix whose elements are conditional variances based on the full GP model. Further, as $\widetilde{\boldsymbol{A}}$ can be provided as a $k{\times}k$ ($k << n$) matrix, and $\widetilde{\boldsymbol{C}}^{-1}$ is sparse, significant reduction in the computational load can be achieved. The spatial process model provided through NNGP can be expressed as follows:

---

[7] Or $m(\boldsymbol{s}_0) + w(\boldsymbol{s}_0)$. (See Cressie, 1993.)
[8] Although the results depend on the ordering of the samples, Datta et al. (2016) showed that NNGP is insensitive to ordering. We performed ordering based on the x-coordinate locations.

$$y \sim N(X\boldsymbol{\beta}, \widetilde{\boldsymbol{\Lambda}}(\tau^2, \boldsymbol{\theta})), \qquad (4)$$

where $\widetilde{\boldsymbol{\Lambda}}(\tau^2, \boldsymbol{\theta}) = \widetilde{\boldsymbol{C}}(\boldsymbol{\theta}) + \tau^2 \boldsymbol{I}$.

The Bayesian Markov chain Monte Carlo (MCMC) (Datta et al., 2016), Hamiltonian Monte Carlo (Wang et al., 2018), and maximum likelihood methods (Saha and Datta, 2018) can be employed to estimate the parameters of the NNGP model. This study employed the MCMC. and as the NNGP parameters are $\boldsymbol{\beta}$ and $\boldsymbol{\phi} = (\tau^2, \sigma^2, \phi)' = (\tau^2, \boldsymbol{\theta})'$, a prior distribution must be set for each parameter and subsequently multiplied by the likelihood function to obtain the conditional posterior distributions (the full Bayesian NNGP). However, because this study employs massive data to a maximum order of $n = 10^6$, implementing the full Bayesian NNGP within a practical computational time is a challenge. Therefore, this study employs the conjugate NNGP, which was proposed by Finley et al. (2017). Assume $\widetilde{\boldsymbol{P}}(\phi)$ is the approximate nearest neighbor of a spatial correlation matrix corresponding to an approximate nearest neighbor of $\widetilde{\boldsymbol{C}}(\boldsymbol{\theta})$. Then, the conjugate NNGP can then be expressed as

$$y \sim N(X\boldsymbol{\beta}, \sigma^2 \widetilde{\boldsymbol{M}}), \qquad (5)$$

where $\widetilde{\boldsymbol{M}} = \widetilde{\boldsymbol{P}}(\phi) + \alpha \boldsymbol{I}$ and $\alpha = \tau^2/\sigma^2$. The reason for employing the conjugate NNGP because, assuming that $\alpha$ and $\phi$ are known, the conjugate normal-inverse Gamma posterior distribution for $\boldsymbol{\beta}$ and $\sigma^2$ can be used. Further, it enables obtaining the predictive distribution for $y(\boldsymbol{s}_0)$ as a $t$-distribution. Thus, performing MCMC sampling is simple. Section 4 explains the setting of the values of $\alpha$ and $\phi$.

### 3.2. ML-based approaches

#### 3.2.1. RF

RF is a bagging-type ensemble of decision trees that trains several trees in parallel. RF was proposed by Breiman (2001) by combining CART and bagging. In the RF algorithm, decision trees constructed from bootstrap samples are combined to conduct a prediction, where each decision tree is trained *independently*. The training procedure can be described as follows: (1) Bootstrap samples are drawn as a randomized subset from the training data, and (2) a decision tree is constructed for every sample, using a randomized subset of predictor variables. This variable selection step helps balance low tree correlation with reasonable predictive strength. (3) Aggregation (i.e., averaging) was performed for each predicted result. Note that there are certain hyperparameters in RF that need to be calibrated, which will be explained in Subsection 4.3.

### 3.2.2. XGBoost

XGBoost is an efficient and scalable approach based on GB developed by Friedman et al. (2000) and Friedman (2001). To generate the final prediction results, GB uses decision trees as weak learners in a *sequential* learning process, in the form of an ensemble of weak predictions such as decision trees. It has three main components: (1) a loss function to be optimized, (2) a weak learner to predict, and (3) an additive model to add weak learners to optimize the loss function. Chen et al. (2016) improved the algorithm by adding a regularization term to reduce overtraining (overfitting) and called it XGBoost. This improved algorithm significantly reduces processing time; however, compared to RF, XGBoost has more hyperparameters that need to be calibrated. Again, this is explained in Subsection 4.3.

### 3.2.3. DNN

A DNN is a mathematical model with a network structure wherein layered units are connected to neighboring layers. Each element that comprises a network is referred to as a unit or node. The first layer is the input layer, while the last is the output layer. The remaining layers referred to as hidden layers. Further, the indices for layers are expressed as $l = 1, \ldots, L$, with the first layer being the input layer and the $L$th the output layer. In a DNN, the previous layer transmits the results of the non-linear transformations on the received inputs to the next layer, which enables the outputs at the output layer to be derived as an estimation result. Thus, an observation was conducted in each layer, via linear transformations using a weight matrix $\boldsymbol{W}_{l+1}$ ($m_l \times m_{(l+1)}$) and non-linear transformations using an activation function $f(.)$. The transformation from the $l$th layer output $z_l$ ($m_l \times 1$) to the $(l+1)$th layer output $z_{l+1}$ ($m_{(l+1)} \times 1$) can be performed using the following equations:

$$\boldsymbol{u}_{l+1} = \boldsymbol{W}_{l+1}\, \boldsymbol{z}_l + \boldsymbol{b}_{l+1}, \tag{6}$$

$$\mathbf{z}_{l+1} = \mathbf{f}(\mathbf{u}_{l+1}),  \quad (7)$$

where $\mathbf{b}_{l+1}$ is the $m_l \times 1$ bias vector and $\mathbf{f}(\mathbf{u}_{l+1})$ is the activation function vector. The final output is denoted by $(z_L \equiv \hat{y})$. When determining $\mathbf{W}_{l+1}$ and $\mathbf{b}_{l+1}$ for a regression (where $y$ is continuous), the mean squared error (MSE) of the actual value $y$ and the predictive value $\hat{y}$ are often used as the loss function $g$, expressed as

$$g = \frac{1}{n}\sum_{i=1}^{n}(y_i - \hat{y}_i)^2. \quad (8)$$

The process of determining $\mathbf{W}_{l+1}$ and $\mathbf{b}_{l+1}$ to minimize $g$ is referred to as DNN learning, which is performed using the gradient algorithm, whereas backpropagation is used to calculate the gradient (LeCun et al., 2015). However, several hyperparameters must be calibrated in the DNN, including the number of layers and units in the hidden layers, learning rate, and batch size. DNN parameter tuning is commonly performed using grid and random searches (Bergstra and Bengio, 2012).

### 3.2.4. ML approaches and spatial dependence

Hengl et al. (2018) argued that RF is essentially a non-spatial approach to spatial prediction because the sampling locations and the general sampling pattern are ignored during the estimation. In this study, we examined and discussed appropriate methods for introducing or considering spatial dependence in the ML framework. In this experiment, we focused on RF because it has fewer hyperparameters than DNN and XGBoost, which may make it easier to understand the impact of differences in model structure on prediction accuracy. Certain possible approaches are as follows:

(1) Geographical covariates

As mentioned in Sekulić et al. (2021), one approach to include a geographic context into RF is to introduce the X and Y coordinates as covariates. We refer to this method as the RF_coordinates. Hengl et al. (2018) proposed the use of buffer distance maps from observation points as covariates. This relative distance method is similar to RF_coordinates; the difference is that the latter has a small number of covariates, while the former has a large number of covariates. In this study, we focused on the RF_coordinates approach.

(2) Spatial autoregressive term

Certain studies have attempted to introduce spatially dependent RF by employing spatial econometrics (Anselin, 1988). Credit (2021) proposed a method for constructing spatially explicit RF models by including spatially lagged (spatial autoregressive) variables to mirror various spatial econometric specifications. The approach entails the introduction of a spatial autoregressive term for the dependent variables (**y**) and explanatory variables (**X**). Sekulić et al. (2021) adopted a similar approach, wherein they directly introduced observations at the $k$ nearest locations and the distances from these locations to the prediction location, which they termed random forest spatial interpolation (RF_si). Note that Sekulić et al. (2021) did not use the weighted average; rather, they directly introduced the actual observed values. In this study, we employ Credit's (2021) approach, that is, the introduction of **Wy** with and without X–Y coordinates (RF_sar and RF_sar_coordinates, respectively), and the RF_si approach by Sekulić et al. (2021).

(3) Eigenvectors of a distance matrix

Considering the results by Murakami and Griffith (2019), Moran eigenvectors from **MCM**, where $M = I - 11'/n$ is a centering operator, **1** is a vector of ones, and **C** is an $n \times n$ spatial weight matrix whose $(i, j)$th element equals $\exp(-d(i,j)/q)$, can be employed to consider spatial dependence. Here, $d(i,j)$ represents the Euclidean distance between the sample sites $i$ and $j$, and $q$ is the maximum length of the minimum spanning tree connecting sample sites (Dray et al., 2006). However, the calculated when $n$ exceeds 1,000 the calculated eigenvalues cannot be guaranteed as accurate (Yamagata and Seya, 2019). Eigen decomposition is possible only when $n < 10,000$ in a standard computing environment. There exist several approximation methods for eigen-decomposition in the ML literature, with a popular approach, called the Nyström extension, being suitable for the Moran eigenvector approximation. Using the results of Murakami and Griffith (2019), the first $h$ approximate eigenpairs can be formulated as follows:

$$\widetilde{E} = [C_{nh} - 1 \otimes (1'_h (C_h + I_h)/h)] E_h (\Lambda_h + I_h), \tag{9}$$

$$\widetilde{\Lambda}_h = \frac{n}{h}(\Lambda_h + I_h) - I_h, \tag{10}$$

where $C_h$ is an $h \times h$ matrix of a spatial weight matrix among $h$ anchor points, defined by $k$-means centers (geometric centers of the clusters defined using the $k$-means method). A greater $h$ yields a better approximation, but results in slower computation; that is, the approximation is influenced by the setting of the number of $h$. Murakami and Griffith (2019) suggested that $h = 200$ be set to balance accuracy and computational efficiency. Although the values should be set only after examining the balance between the sample size and number of hours, this study applies their recommendations.[9] This method is called as the RF_esf and RF_esf_app with and without approximation, respectively. However, the former is applicable only when $n = 10^4$ or less for $C$ constructed is sparse and $MCM$ is not sparse.

(4) Other approaches

Georganos et al. (2019) proposed a method for the remaining algorithms, referred to as geographical random forests, where, for each location $i$, a local RF is computed, but only $k$ number of nearby observations are included. Thus, this results in the calculation of an RF at each training data point, with its own performance, predictive power, and feature importance. Saha et al. (2020) proposed RF-GLS, an extension of RF for dependent error processes similar to the manner in which generalized least squares (GLS) fundamentally extends OLS for linear models under dependence. This extension is based on the equivalent representation of local decision making in a regression tree as a global OLS optimization, which is subsequently replaced by a GLS loss, resulting in a GLS-style regression tree. For spatial settings, RF-GLS coupled with Gaussian process-correlated errors can generate kriging predictions at new locations. However, based on our investigations, these two methods although having potential are not readily applicable to a massive dataset on the order of $n = 10^5$ or higher.

## 4. Empirical comparison

**4.1. Dataset**

The LIFULL HOME data was used on this study to obtain rent price predictions. The dataset used is the same as that used by Seya and Shiroi (2021); however, we repeated the description to maintain consistency. Of 5.33 million properties, 4,588,632 properties (after excluding missing data) were employed as the original data. Further, the original data did not explicitly contain exact property positional coordinates owing to privacy concerns; however, the zip codes were available. Hence, to overcome this issue, the barycentric coordinates for zip codes (X and Y coordinates projected to the UTM54N WGS84 reference system) were employed instead. For cases involving multiple properties sharing the same location (e.g., a different room in the same apartment), small perturbations (random noise) were provided to each positional (X, Y) coordinate within the zip code.[10] The natural logarithm of the monthly rent price (including maintenance fees) in yen is considered the dependent variable,[11] and the explanatory variables used are listed in Tables 1–3. Typical variables were chosen to include descriptors of the location of the condominium (location variables) and the condominium itself (structural variables). "Walk time to nearest (train) station" (m), "Floor-area ratio" (%), and "Use district" (dummies) were employed as the location variables. Meanwhile, "Years built" (month), "Number of rooms" (#), "Direction" (dummies), "Building structure" (dummies), and "Room layout" (dummies) were employed as the structural variables. Further, the number of explanatory variables ($K$) was 43. The descriptive statistics are presented in Tables 1–3.[12]

---
[9] $h = 500$ and $h = 1,000$ have also been tested in our case. We found $h = 200$ to be the optimal setting in terms of predictive accuracy.

[Table 1 Descriptive statistics (continuous variables)].

[Table 2 List of explanatory variables (discrete variables)].

[Table 3 Descriptive statistics (discrete variables)].

## 4.2. Experimental design

For the prediction of the 4,588,632 properties, random sampling was conducted at various sizes ($n = 10^4$, $10^5$, and $10^6$), and 80 % of these data were utilized as training data for the learning models. The remaining 20% were employed as the test (validation) data to assess the predictive accuracy. Thus, the balance of sample sizes for the training and testing data followed three patterns: 8,000 vs. 2,000, 80,000 vs. 20,000, and 800,000 vs. 200,000. Due to the completely random sampling, no containment relations were possible where, for instance, $10^4$ samples were contained in $10^5$ samples. However, as the data size was sufficiently large, a quite low probability exists for the sample bias to conceal any trends. Thus, this study design (based on complete random sampling instead of conditionalization) would not significantly affect the results. Further, OLS, NNGP, RF, and XGBoost were estimated using R language, whereas the DNN was estimated using Python. However, to use the same random numbers for R and Python, a reticulate package that provides an R interface for Python modules, classes, and functions was used.

---

[10] This process can cause certain positional errors; however, as our study is nationwide in scope, we focus only on comparison, these errors can be neglected.

[11] We also considered the linear functional form; however, it resulted in the predictive accuracy at real-scale being worse compared to the log-linear form for all the cases.

[12] Out of all the explanatory variables, information about use district (zoning) and floor-area ratio was often lacking in the original database. Therefore, the National Land Numerical Information database was accessed to create these data separately (http://nlftp.mlit.go.jp/ksj-e/index.html).

Further, for assessing the predictive accuracy, the following error measures were used: the mean absolute error (MAE), root mean squared error (RMSE), and mean absolute percentage error (MAPE). Here, $\hat{y}_m$ and $y_m$ represent the out-of-sample predictive and observed values, respectively, for the $m$th data. However, the first two measures may be affected by outliers for skewed distributions because it is unlikely that the noise will be Gaussian with constant variance. Moreover, calculating the RMSE on a skewed response variable will cause the resulting statistic to be driven primarily by the observations of the highest magnitude (see descriptive statistics). Thus, all error measures were calculated while keeping $y_m$ log-transformed. However, MAPE for log-transformed variables cannot be interpreted as percentages (see Swanson et al., 2000). Hence, we also calculated the MAPE for the real scale.

$$MAE = \frac{1}{M} \sum_{m=1}^{M} |y_m - \hat{y}_m|, \tag{11}$$

$$RMSE = \sqrt{\frac{1}{M} \sum_{m=1}^{M} (y_m - \hat{y}_m)^2}, \tag{12}$$

$$MAPE = \frac{100}{M} \sum_{m=1}^{M} \left|\frac{y_m - \hat{y}_m}{y_m}\right|. \tag{13}$$

### 4.3. Model settings

In this section, we describe the settings of each model, that is, for OLS, NNGP, RF, XGBoost, and DNN. The descriptions for OLS, NNGP, and DNN are similar to those in Seya and Shiroi (2021), but we repeat them to maintain the consistency of this article.

#### 4.3.1. OLS

The variables are presented in Table 1[13]. The rent price was used as the dependent variable. The other variables, except the X and Y coordinates, were used as explanatory variables. Table 4 presents the OLS results for $n = 10^6$. The adjusted $R^2$ value was found to be 0.5165.

[Table 4 Regression analysis results using OLS (example of $n = 10^6$)] near here

#### 4.3.2. NNGP

The full Bayesian NNGP is theoretically sound for estimation and prediction. However, as this study employs massive data, the conjugate NNGP proposed by Finley et al. (2017) was used to aid in the reduction of the computational cost (see Subsection 3.1). The conjugate NNGP allows the acceleration of drawing samples by assuming $\alpha$ and $\phi$ to be known. Finley et al. (2017) proposed assigning values to $\alpha$ and $\phi$ using a grid point search algorithm, which is based on the cross-validation (CV) score. However, the while performing a grid-point search for $n = 10^6$ the computational load it quite high. Therefore, an ad-hoc strategy was adopted in this study while assigning values to $\alpha$ and $\phi$, as detailed by Seya and Shiroi (2021). This was realized by using the spConjNNGP function in the spNNGP package of R. The determination of the number of nearest neighbors for consideration is required when employing the NNGP; thus, based on the CV, the number of nearest neighbors was set to 30 (Seya and Shiroi, 2021).

---

[13] Note that these variables are selected using the step function in R, based on AIC, with $n = 10^4$.

### 4.3.3. RF

In this subsection, we describe the model setup for RF, estimated using the R package ranger (Wright and Ziegler, 2017), which allows for the fast implementation of RF on high-dimensional data. According to Probst et al. (2019), RF has several hyperparameters that must be set by the user. The *number of trees* (number of trees in the forest) must be set sufficiently high, and we set it to 500, which is a typical default value. The *node size* (minimum number of observations in a terminal node) was set to five. This is because it is generally considered to produce good results (Díaz-Uriarte and De Andres, 2006; Goldstein et al., 2011), and a small preliminary experiment showed that the prediction results are fairly robust to these settings. Further, we confirmed that for the hyperparameter *mtry* (number of drawn candidate variables in each split), the default setting of $K/3$, where $K$ is the number of explanatory variables, results in poor performance for some models (RF_si). Therefore, we attempted to optimize the value of *mtry* using the caret package in R to perform a grid search with fivefold cross-validation in the range [3 to $K$]. The calibration results of the *mtry* are shown in Table 5. For RF_sar and RF_sar_coordinates, we employed the R code (rfsi function) provided by the developer,[14] and the number of nearest neighbors $k$ was also cross-validated in the range [3 to 35].

[Table 5 Calibration results of mtry] near here.

### 4.3.4. XGBoost

This subsection describes the model setting for XGBoost, which has a wider range of hyperparameters that need to be calibrated compared to RF. We observed that XGBoost performed worse than (well-tuned) RF when the hyperparameters were set to the default values of xgboost and better than RF when the hyperparameters were calibrated using CV. The hyperparameters in xgboost include *nround*, which controls the maximum number of iterations; *max_depth*, which controls the depth of the tree; *eta*, which controls learning rate; *gamma*, which controls regularization; *colsample_bytree*, which controls the number of features (variables) supplied to a tree; *min_child_weight*, which denotes the minimum number of instances required in a child node; and *subsample*, which is the number of samples (observations) supplied to a tree. Here, *nround* must be set to a sufficiently large value, and we set it to 100, considering the computation time. In addition, we set gamma to 0, indicating no regulation. The others, based on preliminary experiments, were selected from the following ranges via a five-fold cross-validated grid search using the `caret` package in R: *max_depth* = [9, 11, 13], *eta* = [0.1, 0.2], *colsample_bytree* = [0.8, 1], *min_child_weight* = [0.8, 1], and *subsample* = [0.8, 1]. Although these ranges are commonly used, they are suboptimal because the calibrated parameters may provide an outer rather than an inner solution. However, these settings are sufficient to draw the conclusion of interest in this study: XGBoost has a high performance.

---

[14] https://github.com/AleksandarSekulic/RFSI

### 4.3.5. DNN

Several hyperparameters must be determined for DNNs. This study adopted an efficient optimization method called the tree-structured Parzen estimator (TPE) (Bergstra et al., 2011), which can appropriately handle the parameter space of the DNN tree structure, which has been extensively adopted, with its performance being proven to a certain extent (Bergstra et al., 2011; Bergstra et al., 2013). The traditional sigmoid, hyperbolic tangent, softmax, and recently popularized rectified linear unit (ReLU) are a few of the typical activation functions. ReLU offers a computational advantage compared to the others as it induces sparsity in the hidden units (Glorot et al., 2011); furthermore, it accelerates convergence owing to the non-saturation of its gradient (Krizhevsky et al., 2012). Thus, this study adopted ReLU. For the optimizer of the DNN, the results obtained using the typical algorithms, RMSprop (Tieleman and Hinton, 2012), and adaptive moment estimation (Adam) (Kingma and Ba, 2014) have been presented. However, techniques designed to prevent overtraining, such as the introduction of regularized terms and dropouts, have not been employed in this study.

Based on the method described by Seya and Shiroi (2021), the learning procedures were performed as follows. First, considering the *t*th hyperparameter candidate vectors $\boldsymbol{\delta}_t$ coupled with the results of applying five-fold CV with training data for each $\boldsymbol{\delta}_t$ (MSE, Eq. 8), a 50-fold search was performed using the TPE. Second, using the optimal hyperparameter vector and all the training data to assess the predictive accuracy of the testing data a model was created once again. Moreover, the explanatory variables were standardized in advance. We employed Keras[15] for the development of a DNN, while Optuna[16] was used for the implementation of TPE using Python.

### 4.4. Prediction results and discussion

#### 4.4.1. Regression-based versus ML models

The predictive accuracy based on the sample size for each model is illustrated in Figs. 1 (MAE, RMSE, and MAPE for log-scale) and 2 (MAPE for real-scale). For RF, the results for RF_non_spatial and RF_coordinates are shown here. No evident differences were observed in the predictive accuracy of OLD, even if the sample size was increased because OLS does not use local spatial information and thus has a simple model structure such that $n = 10^4$ was sufficiently large to determine the parameters. Moreover, in the case of the real scale, increasing the sample size resulted in the predictive accuracy reducing further. This is because of the increase in the number of high-priced properties in the test data. Seya and Shiroi (2021), who focused on OLS, DNN, and NNGP, showed that NNGP performed the best when considering the three models for all sample sizes and error measures. They concluded that, for rent price prediction models using standard explanatory variables, kriging (NNGP) is useful, provided the sample size is moderate ($n = 10^4$, $10^5$), whereas DNN may be promising if a sufficient sample size is secured ($n = 10^6$).

However, Figs. 1 and 2 show another story. ML models—RF_coordinates and XGBoost—performed considerably better than NNGP, particularly when the sample size was large ($n = 10^6$). In fact, XGBoost achieved the highest prediction accuracy for all sample sizes and error measures for both logarithmic and real scales and for all price bands (when $n = 10^5$, $10^6$) (Fig. 3).[17; 18] According to Fig. 1, the MAE of XGBoost is less than half that of NNGP when $n = 10^6$. These results show that although regression-based approaches have merit in terms of *attribution*, for pure prediction purposes, ML approaches, specifically XGBoost, have an advantage.

---

[15] https://keras.io  
[16] A framework developed via Preferred Networks, Inc. (https://optuna.org)  
[17] We confirmed that, compared to the RF approach, XGBoost is sensitive to hyperparameter settings. When hyperparameters were set to default values given by the R package `xgboost`, XGBoost performed worse than RF.  
[18] We followed the same experimental procedure for the samples taken from the Tokyo metropolitan area and confirmed that a similar conclusion can be obtained. The prediction results are available on request.

[Figure 1. Prediction results by sample size for each model (log-scale):

(a) MAE, (b) RMSE, and (c) MAPE] near here

[Figure 2. Prediction results by sample size for each model (real-scale)] near here

[Figure 3. MAPE per log rent range (in the case of $n = 10^6$)] near here

### 4.4.2. Differences by method for considering the spatial dependence

The predictive accuracy by sample size for each method considering spatial dependence is shown in Fig. 4 (log-scale). It is evident that RF_coordinates performed the best (or at least the second-best) for all sample sizes and error measures. For all methods, except RF_esf_app, considering spatial dependence improved the predictive accuracy. RF_si outperformed the RF_sar. This implies that the weighted average need not be considered when introducing observations at neighboring sites. In fact, the introduction of *Wy* worsens the predictive accuracy for higher- and lower-end markets (Fig. 5). RF_esf_app performed poorly[19], although RF_esf performed better than RF_coordinates for $n = 10^4$. These results may have shown that simply adding spatial coordinates to explanatory variables would be a plausible option to consider spatial dependence in RF.

[Figure 4: Prediction results by sample size for each method of considering spatial dependence (log-scale):

(a) MAE, (b) RMSE, and (c) MAPE] near here

[Figure 5: MAPE per log rent range for each method of considering spatial dependence

(in the case of $n = 10^6$)]

---

[19] Insufficient number of anchor points may be the reason why the accuracy of RF_esf_app is lower than RF_esf.

# 5. Conclusion

The limitations of prior studies, which have the predictive accuracy of real estate sales and rent prices between regression- and ML-based approaches, are the use of small sample sizes and the disregard for spatial dependence, which is an essential characteristic of real estate properties. This study compared and discussed the rent price prediction accuracy of regression- and ML-based approaches by extending the work of Seya and Shiroi (2021) and employing varying sample sizes in a varying order ($n = 10^4$, $10^5$, and $10^6$).

For the regression-based approach, the NNGP model, which enables the application of kriging to large data, was employed. Meanwhile, for the ML-based approaches, certain representative models, namely, XGBoost, RF, and DNN, were employed. To achieve empirical validation, the LIFULL HOME dataset for apartment rent prices in Japan was used in this study. The dataset includes the following variables: rent, lot size, location (municipality, zip code, nearest station, and walk time to the nearest station), year built, room layout, and building structure. Further, the out-of-sample predictive accuracies of the models were compared.

Although Seya and Shiroi (2021) found that NNGP outperformed DNN, particularly when rent prices were around the median, our comparison revealed another story. Our analysis results showed that, with an increase in sample size, the out-of-sample predictive accuracy of XGBoost and RF was higher than that of NNGP. In fact, the performance of XGBoost was the best for all sample sizes. Thus, the results suggest that, although regression-based approaches have merit in terms of *attribution*, ML approaches, specifically XGBoost, have an advantage for pure prediction purposes. We also compared several methods to consider the spatial dependence with RF and found that simply adding spatial coordinates to explanatory variables can be a plausible option.

In future work, it may be important to establish an effective means to set NNGP hyperparameters. Further, it may also be interesting to use other neural network models, including graph convolutional networks. In addition, it is important to conduct experiments with several explanatory variables by using, for instance, pictures that show the floor plans and interior details of each property. Finally, it may be useful to consider spatial dependence when conducting validation (Ploton et al., 2020).

# References

- Abidoye, R. B., & Chan, A. P. (2017). Artificial neural network in property valuation: application framework and research trend. *Property Management*, 35(5), 554–571.

- Anselin, L. (1988). *Spatial Econometrics: Methods and Models*. Dordrecht: Kluwer Academic.

- Banerjee, S., Carlin, B. P., & Gelfand, A. E. (2014). *Hierarchical Modeling and Analysis for Spatial Data* (2nd ed). Boca Raton: Chapman & Hall/CRC.

- Banerjee, S. (2020). Modeling massive spatial datasets using a conjugate Bayesian linear modeling framework. *Spatial Statistics*, 37, 100417.

- Bergstra, J. S., & Bengio, Y. (2012). Random Search for Hyper-Parameter Optimization. *Journal of Machine Learning Research*, 13, 281–305.

- Bergstra, J. S., Bardenet, R., Bengio, Y., & Kégl, B. (2011). Algorithms for Hyper-Parameter Optimization. *Proceedings for Advances in Neural Information Processing Systems*, 24, 2546–2554.

- Bergstra, J. S., Yamins, D., & Cox, D. (2013). Hyperparameter Optimization in Hundreds of Dimensions for Vision Architectures. *Proceedings of the 30th International Conference on Machine Learning*, 28, 115–123.

- Bourassa, S., Cantoni, E., & Hoesli, M. (2010). Predicting House Prices with Spatial Dependence: A Comparison of Alternative Methods. *Journal of Real Estate Research*, 32(2), 139–159.

- Breiman, L. (1996). Bagging predictors. *Machine learning*, 24(2), 123–140.

- Breiman, L. (2001). Random forests. *Machine learning*, 45(1), 5–32.

- Breiman, L., Friedman, J., Stone, C. J., & Olshen, R. A. (1984). *Classification and regression trees*. Boca Raton: Routledge.

- Brunsdon, C., Fotheringham, S., & Charlton, M. (1998). Geographically Weighted Regression. *Journal of the Royal Statistical Society: Series D (The Statistician)*, 47(3), 431–443.

- Čeh, M., Kilibarda, M., Lisec, A., & Bajat, B. (2018). Estimating the Performance of Random Forest Versus Multiple Regression for Predicting Prices of the Apartments. *ISPRS International*

*Journal of Geo-Information*, 7(5), 168.

- Chen, T., & Guestrin, C. (2016). XGBoost: A scalable tree boosting system. *Proceedings of the 22nd ACM SIGKDD International Conference on Knowledge Discovery and Data Mining*, 785–794.

- Cressie, N. (1993). *Statistics for Spatial Data*. New York: Wiley.

- Cressie, N., & Wikle, C. K. (2011). *Statistics for Spatio-Temporal Data*. Hoboken: John Wiley and Sons.

- Credit, K. (2021). Spatial models or random forest? Evaluating the use of spatially explicit machine learning methods to predict employment density around new transit stations in Los Angeles. *Geographical Analysis*, in press.

- Datta, A., Banerjee, S., Finley, A. O., & Gelfand, A. E. (2016). Hierarchical Nearest-Neighbor Gaussian Process Models for Large Geostatistical Datasets. *Journal of the American Statistical Association*, 111(514), 800–812.

- Dambon, J. A., Sigrist, F., & Furrer, R. (2021). Maximum likelihood estimation of spatially varying coefficient models for large data with an application to real estate price prediction. *Spatial Statistics*, 41, 100470.

- Díaz-Uriarte, R., & De Andres, S. A. (2006). Gene selection and classification of microarray data using random forest. *BMC Bioinformatics*, 7, 3.

- Dubin, R. A. (1988). Estimation of regression coefficient in the presence of spatially autocorrelated error terms. *The Review of Economics and Statistics*, 70(3), 466–474.

- Dray, S., Legendre, P., & Peres-Neto, P. R. (2006). Spatial modelling: a comprehensive framework for principal coordinate analysis of neighbour matrices (PCNM). *Ecological modelling*, 196(3–4), 483–493.

- Efron, B. (2020). Prediction, Estimation, and Attribution. *Journal of the American Statistical Association*, 115(530), 636–655.

- Finley, A. O., Datta, A., Cook, B. C., Morton, D. C., Andersen, H. E., & Banerjee, S. (2017). Applying nearest neighbor gaussian processes to massive spatial data sets: Forest canopy height

- prediction across tanana valley alaska. *arXiv preprint*, 1702.00434.

- Friedman, J. H. (2001). Greedy function approximation: a gradient boosting machine. *Annals of statistics*, 29(5), 1189–1232.

- Friedman, J., Hastie, T., & Tibshirani, R. (2000). Additive logistic regression: a statistical view of boosting (with discussion and a rejoinder by the authors). *Annals of statistics*, 28(2), 337–407.

- Gelfand, A. E., Kim, H. J., Sirmans, C. F., & Banerjee, S. (2003). Spatial modeling with spatially varying coefficient processes. *Journal of the American Statistical Association*, 98(462), 387–396.

- Georganos, S., Grippa, T., Niang Gadiaga, A., Linard, C., Lennert, M., Vanhuysse, S., Mboga, N., Wolff, E., & Kalogirou, S. (2019). Geographical random forests: a spatial extension of the random forest algorithm to address spatial heterogeneity in remote sensing and population modelling. *Geocarto International*, 36(2), 121–136.

- Glorot, X., Bordes, A., & Bengio, Y. (2011). Deep Sparse Rectifier Neural Networks. *Proceedings of the Fourteenth International Conference on Artificial Intelligence and Statistics*, 15, 315–323.

- Goldstein, B. A., Polley, E. C., & Briggs, F. B. (2011). Random forests for genetic association studies. *Statistical Applications in Genetics and Molecular Biology*, 10(1), 32.

- Hayunga, D. K., & Pace, R. K. (2010). Spatial statistics applied to commercial real estate. *The Journal of Real Estate Finance and Economics*, 41(2), 103–125.

- Heaton, M. J., Datta, A., Finley, A. O., Furrer, R., Guinness, J., Guhaniyogi, R., Gerber, F., Gramacy, R. B., Hammerling, D., Katzfuss, M., Lindgren, F., Nychka, D. W., Sun, F., & Zammit-Mangion, A. (2019). A case study competition among methods for analyzing large spatial data. *Journal of Agricultural, Biological and Environmental Statistics*, 24(3), 398–425.

- Hengl, T., Nussbaum, M., Wright, M. N., Heuvelink, G. B., & Gräler, B. (2018). Random forest as a generic framework for predictive modeling of spatial and spatio-temporal variables. *PeerJ*, 6, e5518.

- Ho, W. K., Tang, B. S., & Wong, S. W. (2021). Predicting property prices with machine learning algorithms. *Journal of Property Research*, 38(1), 48–70.

- Iranzad, R., Liu, X., Chaovalitwongse, W., Hippe, D. S., Wang, S., Han, J., Thammasorn, P., Duan, C., Zeng, J., & Bowen, S. R. (2021). Boost-S: Gradient Boosted Trees for Spatial Data and Its Application to FDG-PET Imaging Data. *arXiv preprint*, 2101.11190.

- James, V., Wu, S., Gelfand, A., & Sirmans, C. (2005). Apartment rent prediction using spatial modeling. *Journal of Real Estate Research*, 27(1), 105–136.

- Krizhevsky, A., Sutskever, I., & Hinton, G. E. (2012). Imagenet classification with deep convolutional neural networks. *Proceedings of the 25th International Conference on Neural Information Processing Systems*, 1, 1097–1105.

- Kingma, D. P., & Ba, J. (2014). Adam: A method for stochastic optimization. *arXiv preprint*, 1412.6980.

- LeCun, Y., Bengio, Y., & Hinton, G. (2015). Deep learning. *Nature*, 521(7553), 436–444.

- Li, Z., & Fotheringham, A. S. (2020). Computational improvements to multi-scale geographically weighted regression. *International Journal of Geographical Information Science*, 34(7), 1378–1397.

- Mayer, M., Bourassa, S. C., Hoesli, M., & Scognamiglio, D. (2019). Estimation and updating methods for hedonic valuation. *Journal of European Real Estate Research*, 12(1), 134–150.

- Ming, Y., Zhang, J., Qi, J., Liao, T., Wang, M., & Zhang, L. (2020). Prediction and Analysis of Chengdu Housing Rent Based on XGBoost Algorithm. Proceedings of the 2020 3rd International Conference on Big Data Technologies, 1–5.

- Murakami, D., & Griffith, D. A. (2019). Eigenvector spatial filtering for large data sets: fixed and random effects approaches. *Geographical Analysis*, 51(1), 23–49.

- Murakami, D., Tsutsumida, N., Yoshida, T., Nakaya, T., & Lu, B. (2020). Scalable GWR: A linear-time algorithm for large-scale geographically weighted regression with polynomial kernels. *Annals of the American Association of Geographers*, 111(2), 459–480.

- Murakami, D., Yoshida, T., Seya, H., Griffith, D. A., & Yamagata, Y. (2017). A Moran coefficient-based mixed effects approach to investigate spatially varying relationships. *Spatial Statistics*, 19, 68–89.

- Pace, R. K., & LeSage, J. P. (2004). Spatial statistics and real estate. *The Journal of Real Estate Finance and Economics*, 29(2), 147–148.

- Pace, R. K., & Hayunga, D. (2020). Examining the Information Content of Residuals from Hedonic and Spatial Models Using Trees and Forests. *The Journal of Real Estate Finance and Economics*, 60(1), 170–180.

- Ploton, P., Mortier, F., Réjou-Méchain, M., Barbier, N., Picard, N., Rossi, V., Dormann, C., Cornu, G., Viennois, G., Bayol, N., Lyapustin, A., Gourlet-Fleury, S., & Pélissier, R. (2020). Spatial validation reveals poor predictive performance of large-scale ecological mapping models. *Nature Communications*, 11(1), 4540.

- Probst, P., Wright, M. N., & Boulesteix, A. L. (2019). Hyperparameters and tuning strategies for random forest. *Wiley Interdisciplinary Reviews: Data Mining and Knowledge Discovery*, 9(3), e1301.

- Saha, A., & Datta, A. (2018). BRISC: bootstrap for rapid inference on spatial covariances. *Stat*, 7(1), e184.

- Saha, A., Basu, S., & Datta, A. (2020). Random Forests for dependent data. *arXiv preprint*, 2007.15421.

- Sekulić, A., Kilibarda, M., Protić, D., & Bajat, B. (2021). A high-resolution daily gridded meteorological dataset for Serbia made by Random Forest Spatial Interpolation. *Scientific Data*, 8, 123.

- Seya, H., & Shiroi, D. (2021). A comparison of residential apartment rent price predictions using a large data set: Kriging versus Deep Neural Network. *Geographical Analysis*, in press.

- Seya, H., Tsutsumi, M., Yoshida, Y., & Kawaguchi, Y. (2011). Empirical comparison of the various spatial prediction models: In spatial econometrics, spatial statistics, and semiparametric statistics. *Procedia-Social and Behavioral Sciences*, 21, 120–129.

- Swanson, D. A., Tayman, J., & Barr, C. F. (2000). A note on the measurement of accuracy for subnational demographic estimates. *Demography*, 37(2), 193–201.

- Tieleman, T., & Hinton, G. (2012). Lecture 6.5-RMSprop: Divide the gradient by a running

average of its recent magnitude. *COURSERA: Neural Networks for Machine Learning*, 4(2), 26–31.

- Tsutsumi, M., & Seya, H. (2009). Hedonic Approaches Based on Spatial Econometrics and Spatial Statistics: Application to Evaluation of Project Benefits. *Journal of Geographical Systems*, 11(4), 357–380.
- Valier, A. (2020). Who performs better? AVMs vs hedonic models. *Journal of Property Investment & Finance*, 38(3), 213–225.
- Vecchia, A. V. (1988). Estimation and Model Identification for Continuous Spatial Processes. *Journal of the Royal Statistical Society: Series B (Methodological)*, 50(2), 297–312.
- Wang, C., Puhan, M. A., Furrer, R., & Group, S. S. (2018). Generalized spatial fusion model framework for joint analysis of point and areal data. *Spatial Statistics*, 23, 72–90.
- Wright, M.N., & Ziegler, A. (2017). ranger: A fast implementation of random forests for high dimensional data in C++ and R. *Journal of Statistical Software*, 77(1), 1–17.
- Yamagata, Y., & Seya, H. (2019). *Spatial Analysis Using Big Data: Methods and Urban Applications*. Cambridge, USA: Academic Pres.
- Zhang, L., Datta, A., & Banerjee, S. (2019). Practical Bayesian modeling and inference for massive spatial data sets on modest computing environments. *Statistical Analysis and Data Mining: The ASA Data Science Journal*, 12(3), 197–209.
- Zurada, J., Levitan, A., & Guan, J. (2011). A comparison of regression and artificial intelligence methods in a mass appraisal context. *Journal of Real Estate Research*, 33(3), 349–388.

# Tables

**Table 1: Descriptive statistics (continuous variables)**

|  | Min | Max | Median | Mean | SD |
|---|---|---|---|---|---|
| Rent price (yen) | 5250 | 1250000000 | 63000 | 72850 | 1381893 |
| Years built (month) | 5 | 1812 | 228 | 236 | 135.6 |
| Walk time to nearest (train) station (m) | 1 | 88000 | 640 | 781.5 | 661.3 |
| Number of rooms (#) | 1 | 50 | 1 | 1.48 | 0.71 |
| Floor-area ratio (%) | 50 | 1000 | 200 | 234.1 | 130.6 |
| X (km) | −841 | 783.1 | 352.2 | 181.5 | 273.3 |
| Y (km) | 2958 | 5029 | 3931 | 3942 | 195.3 |

The "rent price" includes maintenance fees

**Table 2: List of explanatory variables (discrete variables)**

| Direction | North, Northeast, East, Southeast, South, Southwest, West, Northwest, Other |
|---|---|
| Building structure | W, B, S, RC, SRC, PC, HPC, LS, ALC, RCB, Others |
| Room layout | R, K, SK, DK, SDK, LK, SLK, LDK, SLDK |
| Use district | Category I exclusively low residential zone (1 Exc Low), Category II exclusively low residential zone (2 Exc Low), Category I exclusively high-medium residential zone (1 Exc Med), Category II exclusively high-medium residential zone (2 Exc Med), Category I residential zone (1 Res), Category II residential zone (2 Res), Quasi-residential zone (Quasi-Res), Neighborhood commercial zone (Neighborhood Comm), Commercial zone (Commercial), Quasi-Industrial zone (Quasi-Ind), Industrial zone (Industrial), Exclusive industrial zone (Exc Ind), Others (Others) |

For building structure: W: Wooden; B: Concrete block; S: Steel frame; RC: Reinforced concrete; SRC: Steel frame reinforced concrete; PC: precast concrete; HPC: Hard precast concrete; LS: Light steel, RCB: Reinforced concrete block

For room layout: The R refers to a room where there is only one room and there is no wall to separate the bedroom from the kitchen. For the others, K: includes a kitchen; D: includes a dining room: L: includes a living room; S: additional storage room. For example, LDK is a Living, Dining, and Kitchen area.

For use district: Category I exclusively low residential zone, Category II exclusively low residential zone, Category I exclusively medium-high residential zone, Category II exclusively medium-high residential zone, Category I residential zone, Category II residential zone, Quasi-residential zone, Neighborhood commercial zone, Commercial zone, Quasi-industrial zone, Industrial zone, Exclusively industrial zone

**Table 3: Descriptive statistics (discrete variables)**

| Direction | Count | Share | Structure | Count | Share | Use district | Count | Share | Room layout | Count | Share |
|---|---|---|---|---|---|---|---|---|---|---|---|
| North | 156843 | 0.0342 | W | 1024081 | 0.2232 | 1 Exc Low | 780638 | 0.1701 | R | 423815 | 0.0924 |
| Northeast | 81173 | 0.0177 | B | 570 | 0.0001 | 2 Exc Low | 25793 | 0.0056 | K | 1729903 | 0.3770 |
| East | 595252 | 0.1297 | S | 844184 | 0.1840 | 1 Exc Med | 689879 | 0.1503 | SK | 6919 | 0.0015 |
| Southeast | 473041 | 0.1031 | RC | 1892428 | 0.4124 | 2 Exc Med | 321441 | 0.0701 | DK | 890584 | 0.1941 |
| South | 1749315 | 0.3812 | SRC | 190048 | 0.0414 | 1 Res | 1030319 | 0.2245 | SDK | 5123 | 0.0011 |
| Southwest | 458125 | 0.0998 | PC | 11924 | 0.0026 | 2 Res | 211076 | 0.0460 | LK | 516 | 0.0001 |
| West | 404994 | 0.0883 | HPS | 802 | 0.0002 | Quasi-Res | 59863 | 0.0130 | SLK | 138 | 0.0000 |
| Northwest | 78836 | 0.0172 | LS | 559974 | 0.1220 | Neighborhood Comm | 386531 | 0.0842 | LDK | 1505821 | 0.3282 |
| Others | 591053 | 0.1288 | ALC | 58373 | 0.0127 | Commercial | 615630 | 0.1342 | SLDK | 25813 | 0.0056 |
| | | | RCB | 597 | 0.0001 | Quasi-Ind | 371672 | 0.0810 | | | |
| | | | Others | 5651 | 0.0012 | Industrial | 83826 | 0.0183 | | | |
| | | | | | | Exc Ind | 11949 | 0.0026 | | | |
| | | | | | | Others | 15 | 0.0000 | | | |

Note: # denotes the number of cases

**Table 4: Regression analysis results using OLS (example of $n = 10^6$)**

| Variables | Estimate | $t$-values | |
|---|---|---|---|
| Constant term | $1.08 \times 10^1$ | $4.50 \times 10^3$ | *** |
| Years built | $-1.15 \times 10^{-3}$ | $-4.42 \times 10^2$ | *** |
| Walk time to nearest station | $-4.88 \times 10^{-5}$ | $-9.87 \times 10^1$ | *** |
| Floor-area ratio | $1.30 \times 10^{-3}$ | $2.30 \times 10^2$ | *** |
| Number of rooms | $1.49 \times 10^{-1}$ | $2.57 \times 10^2$ | *** |
| Direction_Northeast | $8.09 \times 10^{-2}$ | $2.80 \times 10^1$ | *** |
| Direction_East | $-4.45 \times 10^{-3}$ | $-2.32 \times 10^0$ | * |
| Direction_Southeast | $5.40 \times 10^{-3}$ | $2.72 \times 10^0$ | ** |
| Direction_South | $-2.33 \times 10^{-2}$ | $-1.29 \times 10^1$ | *** |
| Direction_Southwest | $2.46 \times 10^{-3}$ | $1.23 \times 10^0$ | |
| Direction_West | $1.94 \times 10^{-3}$ | $9.67 \times 10^0$ | *** |
| Direction_Northwest | $7.39 \times 10^{-2}$ | $2.53 \times 10^1$ | *** |
| Direction_Others | $-6.85 \times 10^{-2}$ | $-3.53 \times 10^1$ | *** |
| Structure_B | $1.88 \times 10^{-1}$ | $6.39 \times 10^0$ | *** |
| Structure_S | $9.41 \times 10^{-2}$ | $9.39 \times 10^1$ | *** |
| Structure_RC | $2.40 \times 10^{-1}$ | $2.71 \times 10^2$ | *** |
| Structure_SRC | $3.67 \times 10^{-1}$ | $2.06 \times 10^2$ | *** |
| Structure_PC | $2.14 \times 10^{-1}$ | $3.48 \times 10^1$ | *** |
| Structure_HPC | $9.13 \times 10^{-2}$ | $4.19 \times 10^0$ | *** |
| Structure_LS | $5.34 \times 10^{-2}$ | $4.75 \times 10^1$ | *** |
| Structure_ALC | $9.17 \times 10^{-2}$ | $3.19 \times 10^1$ | *** |
| Structure_RCB | $1.20 \times 10^{-1}$ | $4.55 \times 10^0$ | *** |
| Structure_Others | $1.61 \times 10^{-1}$ | $1.81 \times 10^1$ | *** |
| Room layout_K | $4.22 \times 10^{-2}$ | $3.62 \times 10^1$ | *** |
| Room layout_SK | $1.10 \times 10^{-1}$ | $1.39 \times 10^1$ | *** |
| Room layout_DK | $1.37 \times 10^{-1}$ | $1.00 \times 10^2$ | *** |
| Room layout_SDK | $3.65 \times 10^{-1}$ | $3.95 \times 10^1$ | *** |
| Room layout_LK | $2.79 \times 10^{-1}$ | $1.01 \times 10^1$ | *** |
| Room layout_SLK | $3.04 \times 10^{-1}$ | $5.75 \times 10^0$ | *** |
| Room layout_LDK | $2.76 \times 10^{-1}$ | $2.12 \times 10^2$ | *** |
| Room layout_SLDK | $6.06 \times 10^{-1}$ | $1.39 \times 10^2$ | *** |
| Use district_2 Exc Low | $-1.15 \times 10^{-1}$ | $-2.72 \times 10^1$ | *** |
| Use district_1 Exc Med | $-1.52 \times 10^{-1}$ | $-1.22 \times 10^2$ | *** |
| Use district_2 Exc Med | $-2.77 \times 10^{-1}$ | $-1.83 \times 10^2$ | *** |
| Use district_1 Res | $-2.36 \times 10^{-1}$ | $-1.97 \times 10^2$ | *** |
| Use district_2 Res | $-2.48 \times 10^{-1}$ | $-1.38 \times 10^2$ | *** |
| Use district_ Quasi-Res | $-2.92 \times 10^{-1}$ | $-9.98 \times 10^1$ | *** |
| Use district_ Neighborhood Comm | $-2.63 \times 10^{-1}$ | $-1.56 \times 10^2$ | *** |
| Use district_ Commercial | $-4.63 \times 10^{-1}$ | $-1.83 \times 10^2$ | *** |
| Use district_ Quasi-Ind | $-1.91 \times 10^{-1}$ | $-1.26 \times 10^2$ | *** |
| Use district_ Industrial | $-2.46 \times 10^{-1}$ | $-9.76 \times 10^1$ | *** |
| Use district_ Exc Ind | $-3.17 \times 10^{-1}$ | $-5.13 \times 10^1$ | *** |
| Use district_Others | $5.00 \times 10^{-1}$ | $3.57 \times 10^0$ | *** |
| Adjusted $R^2$ | | 0.5165 | |

* significant at 5%; ** significant at 1%; *** significant at 0.1%.

Table 5: Calibration results of mtry

| mtry | RF_esf | RF_esf_app | RF_sar_coordinates | RF_sar | RF_si | RF_coordinates | RF_non_spatial |
|---|---|---|---|---|---|---|---|
| $n = 10^4$ | 7 | 7 | 3 | 3 | 29 | 5 | 5 |
| $n = 10^5$ | - | 9 | 3 | 3 | 51 | 5 | 5 |
| $n = 10^6$ | - | 13 | 5 | 5 | 58 | 7 | 7 |

# Figure captions

- Figure 1: Prediction results by sample size for each model (log-scale): (a) MAE, (b) RMSE, and (c) MAPE
- Figure 2: Prediction results by sample size for each model (real-scale)
- Figure 3: MAPE per log rent range (in the case of $n = 10^6$)
- Figure 4: Prediction results by sample size for each method of considering spatial dependence (log-scale): (a) MAE, (b) RMSE, and (c) MAPE
- Figure 5: MAPE per log rent range for each method of considering spatial dependence (in the case of $n = 10^6$)

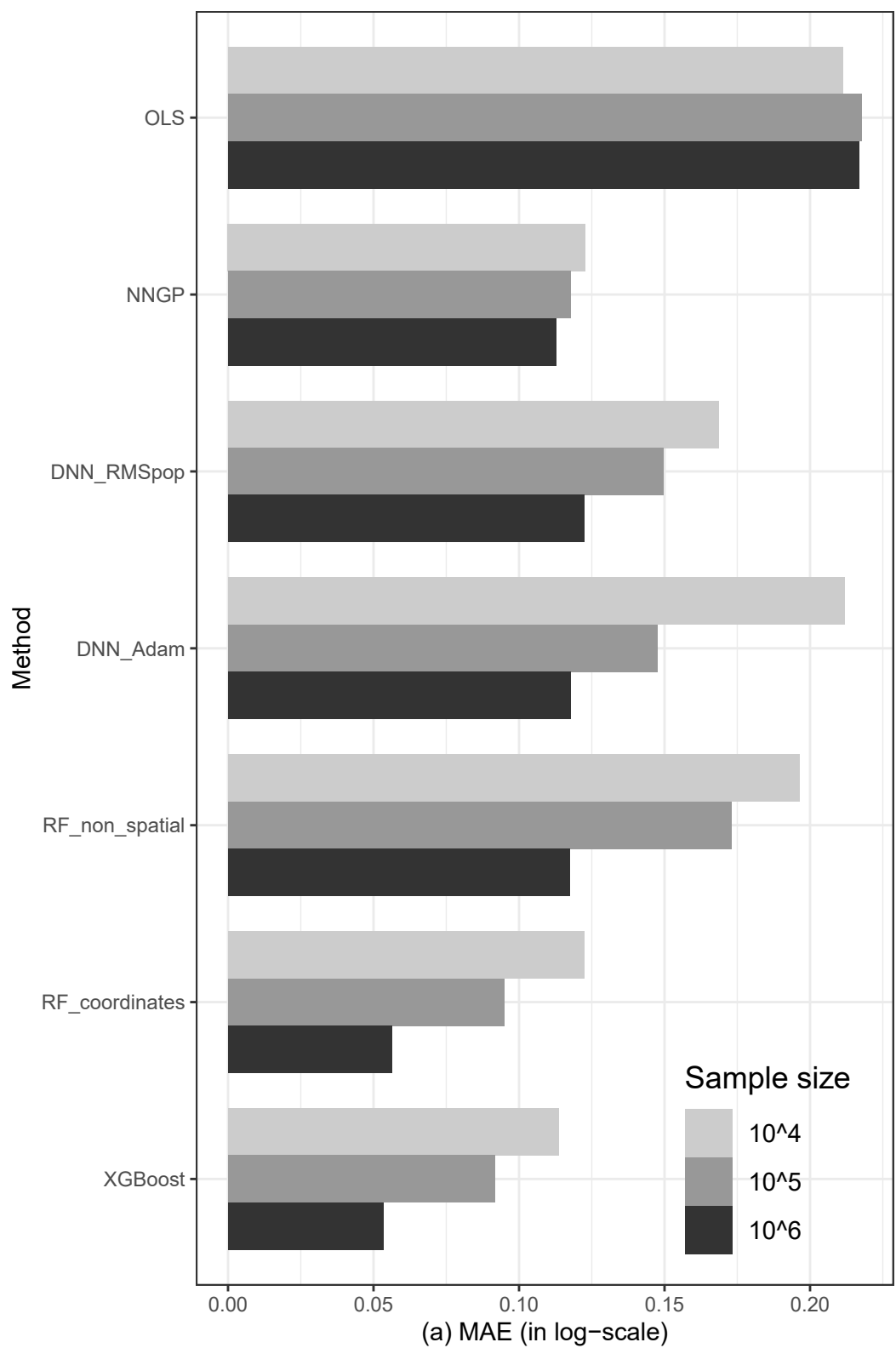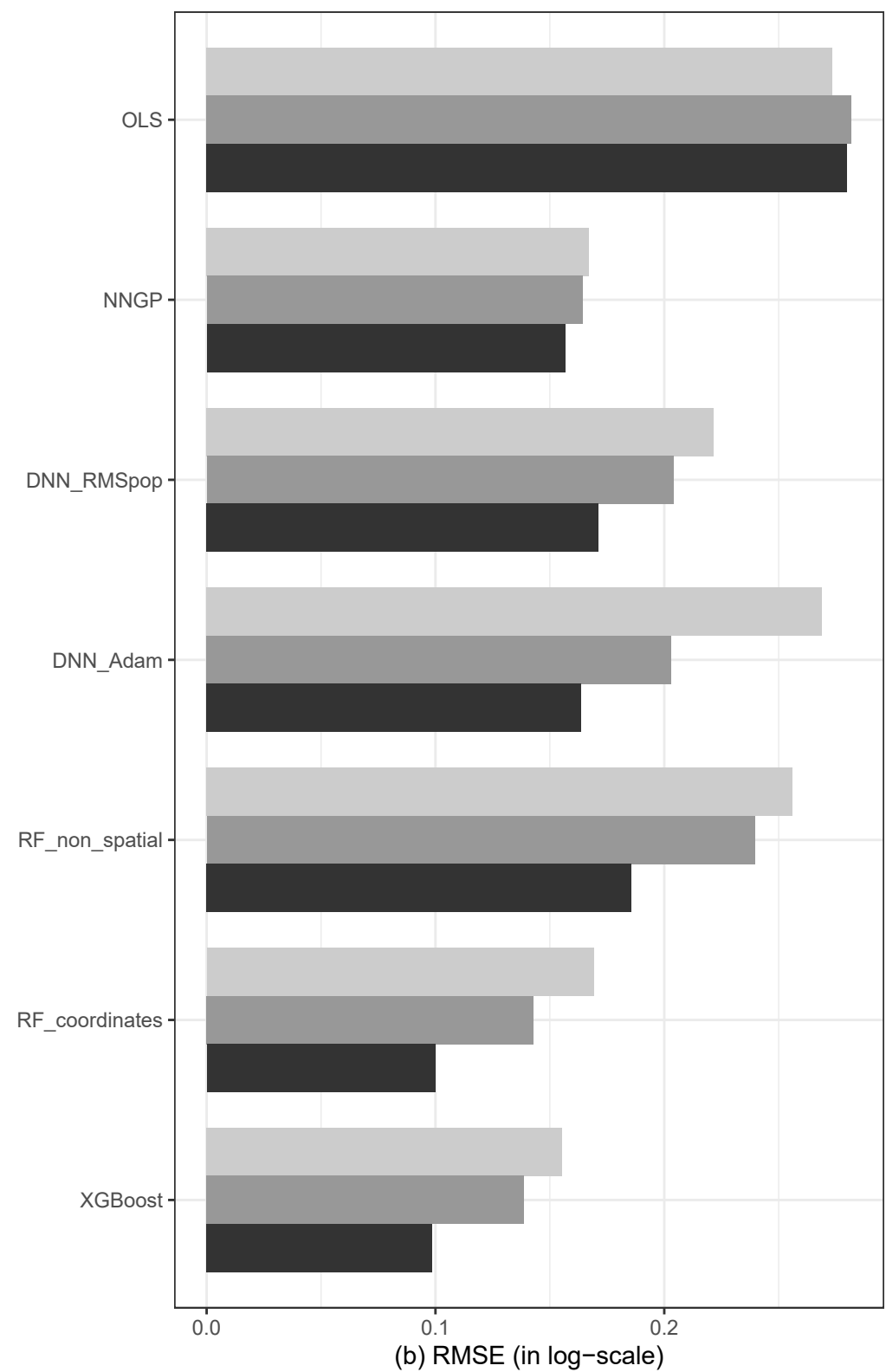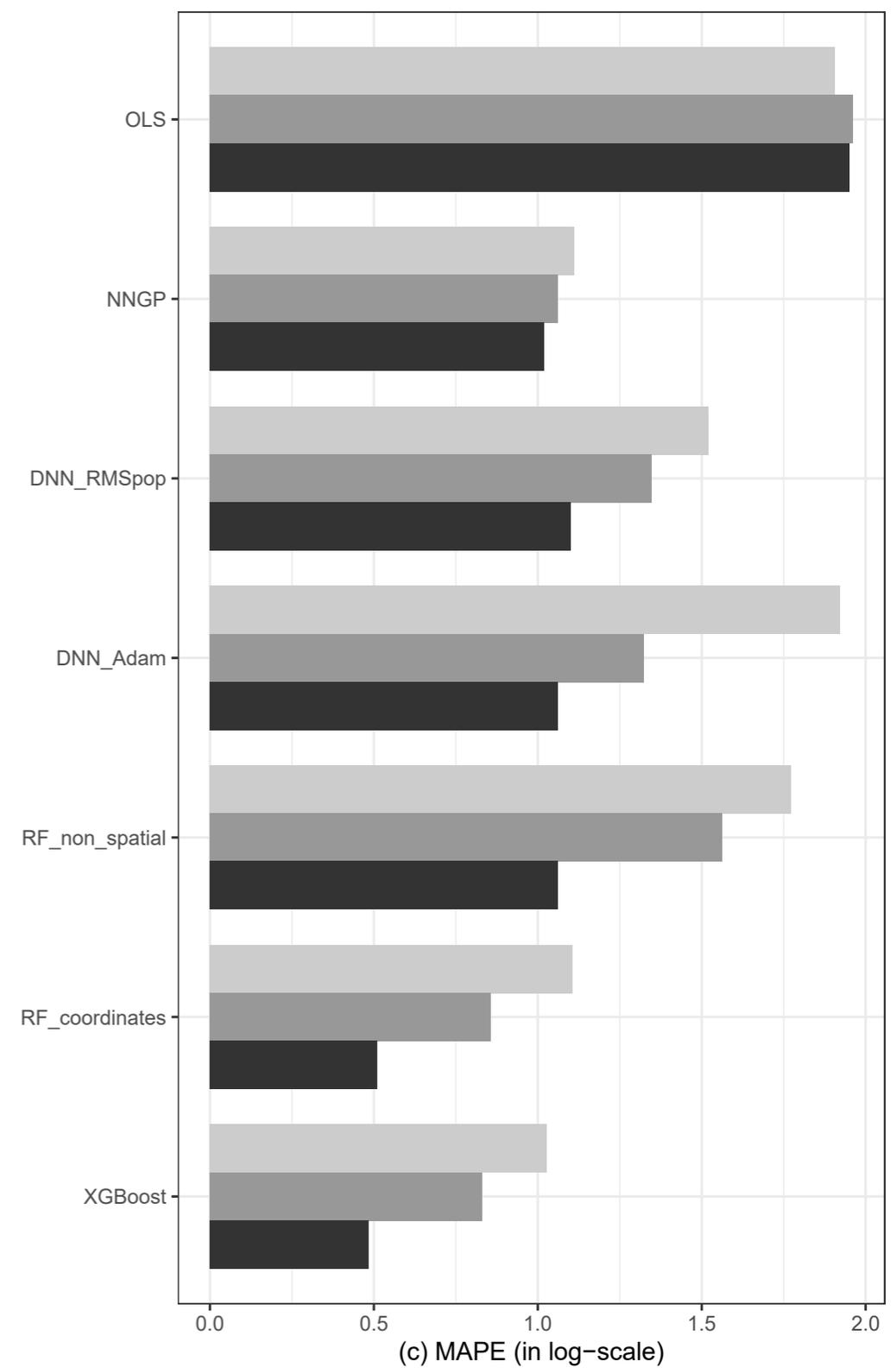

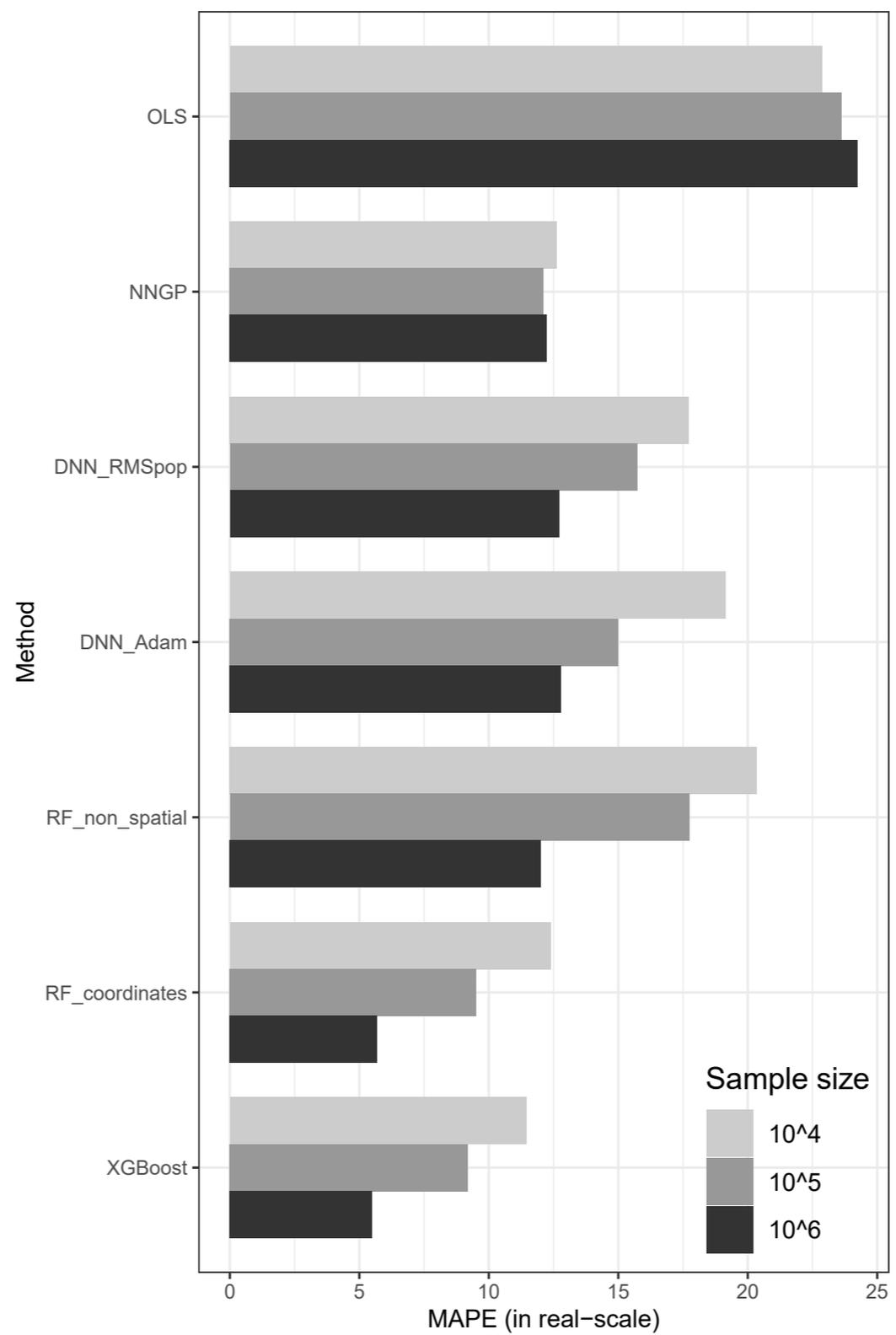

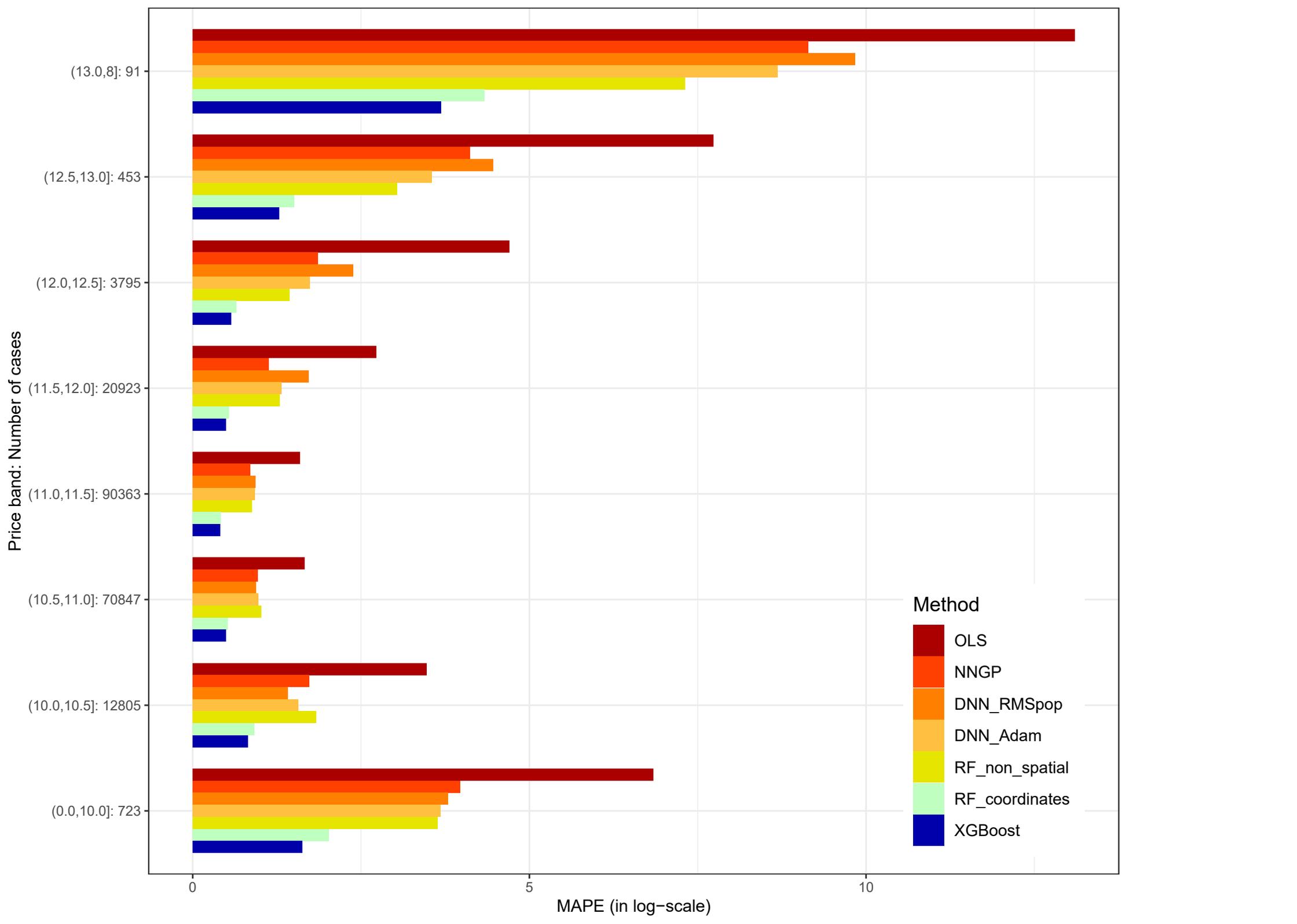

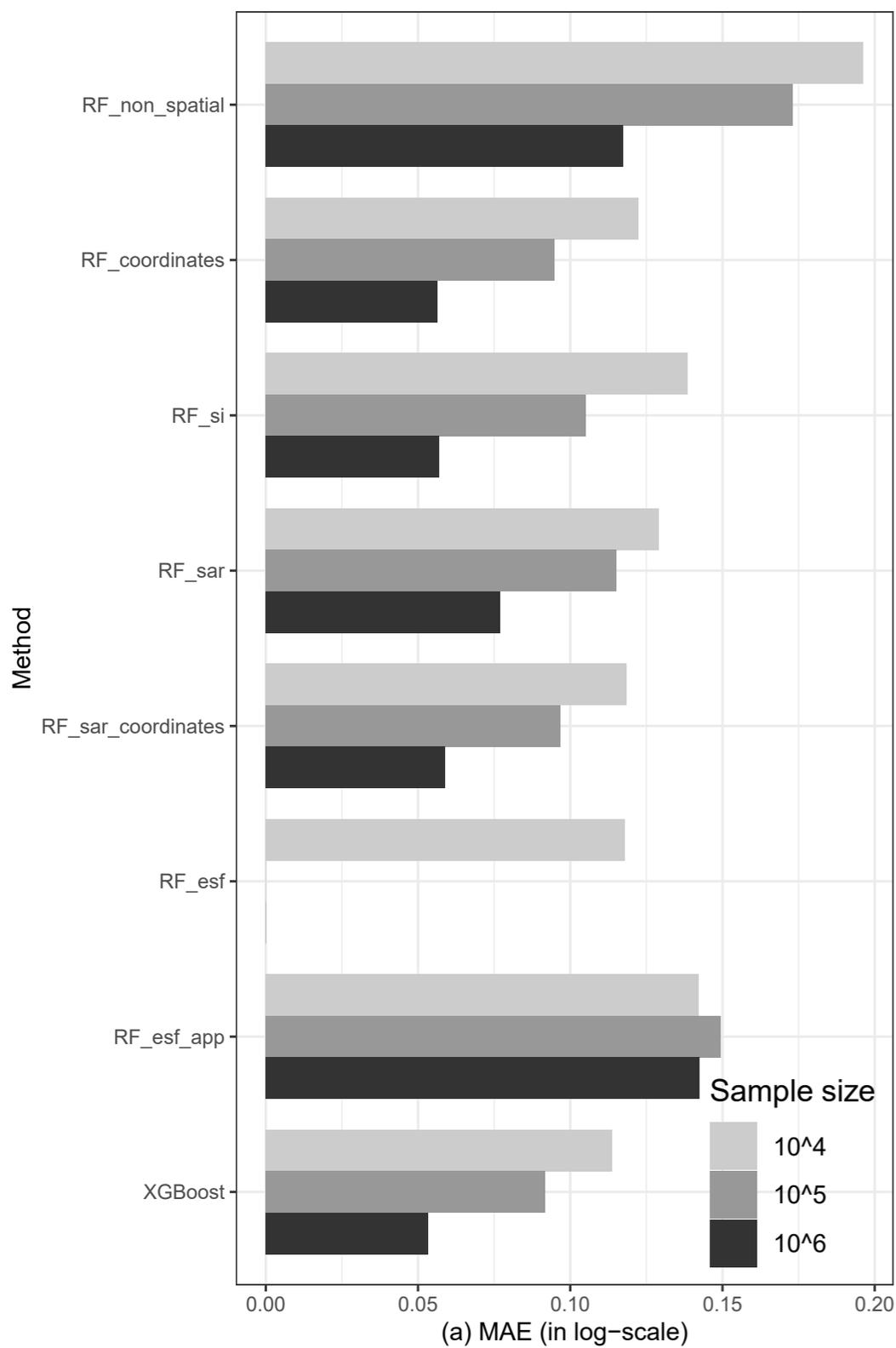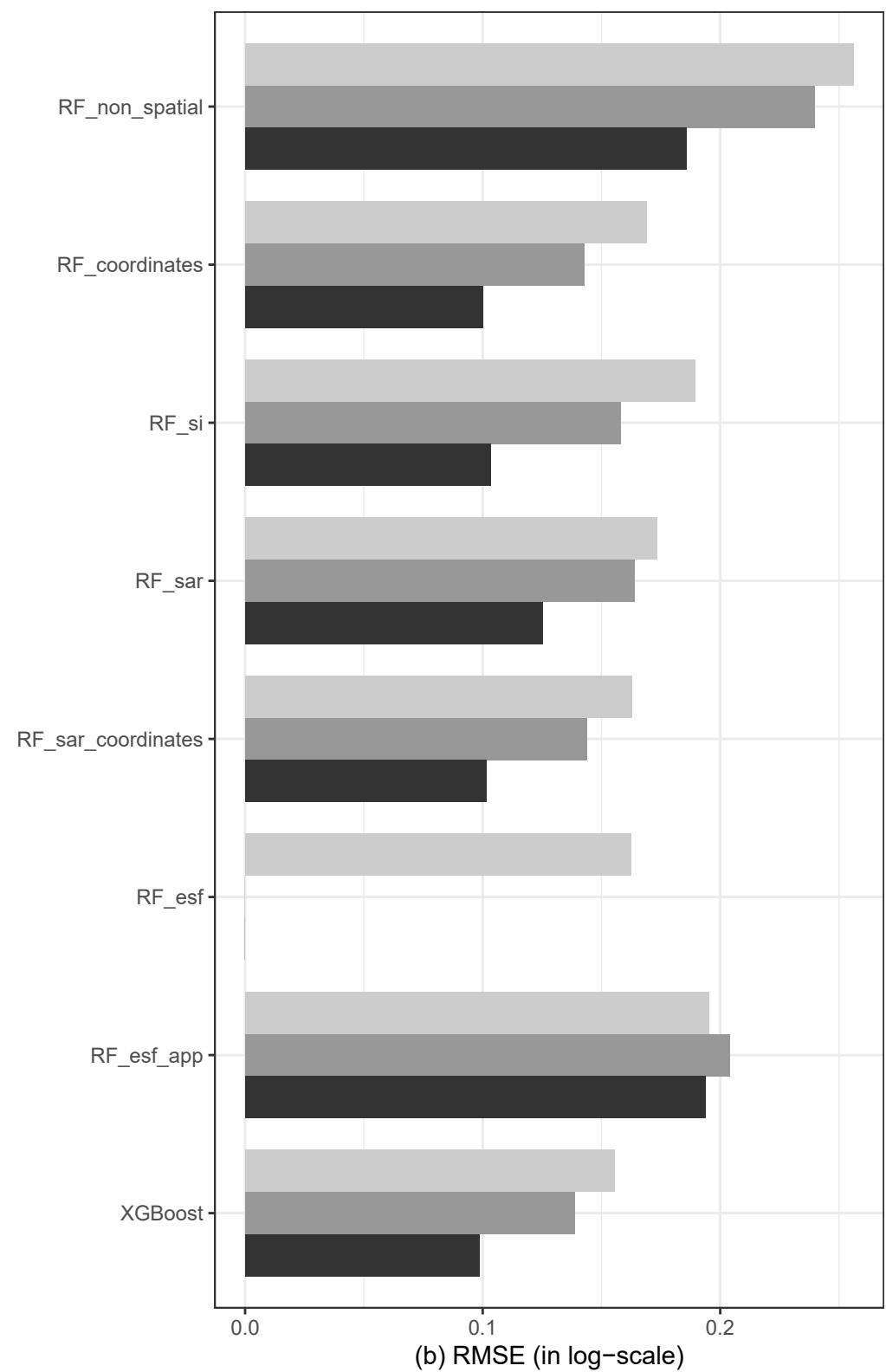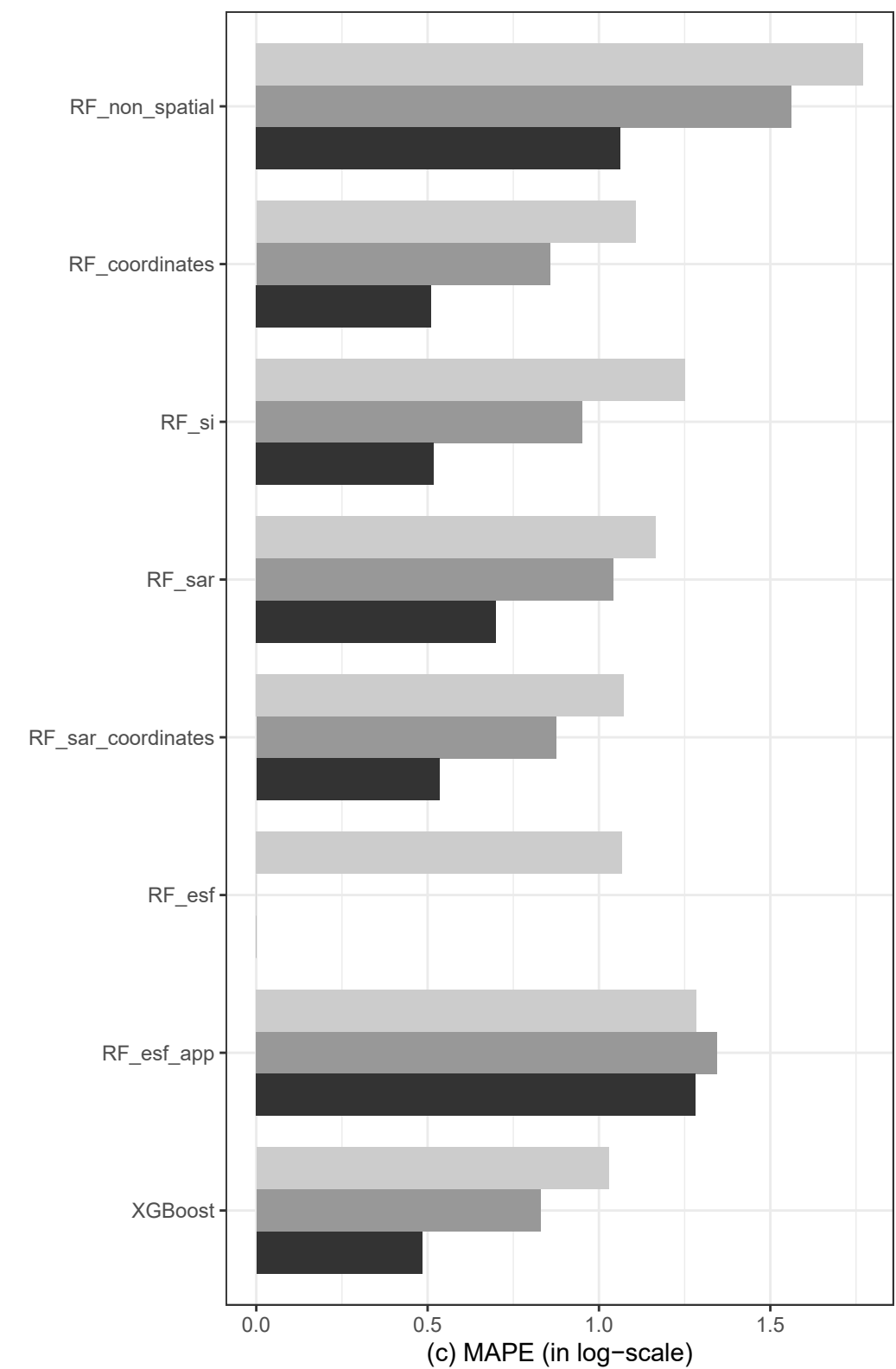

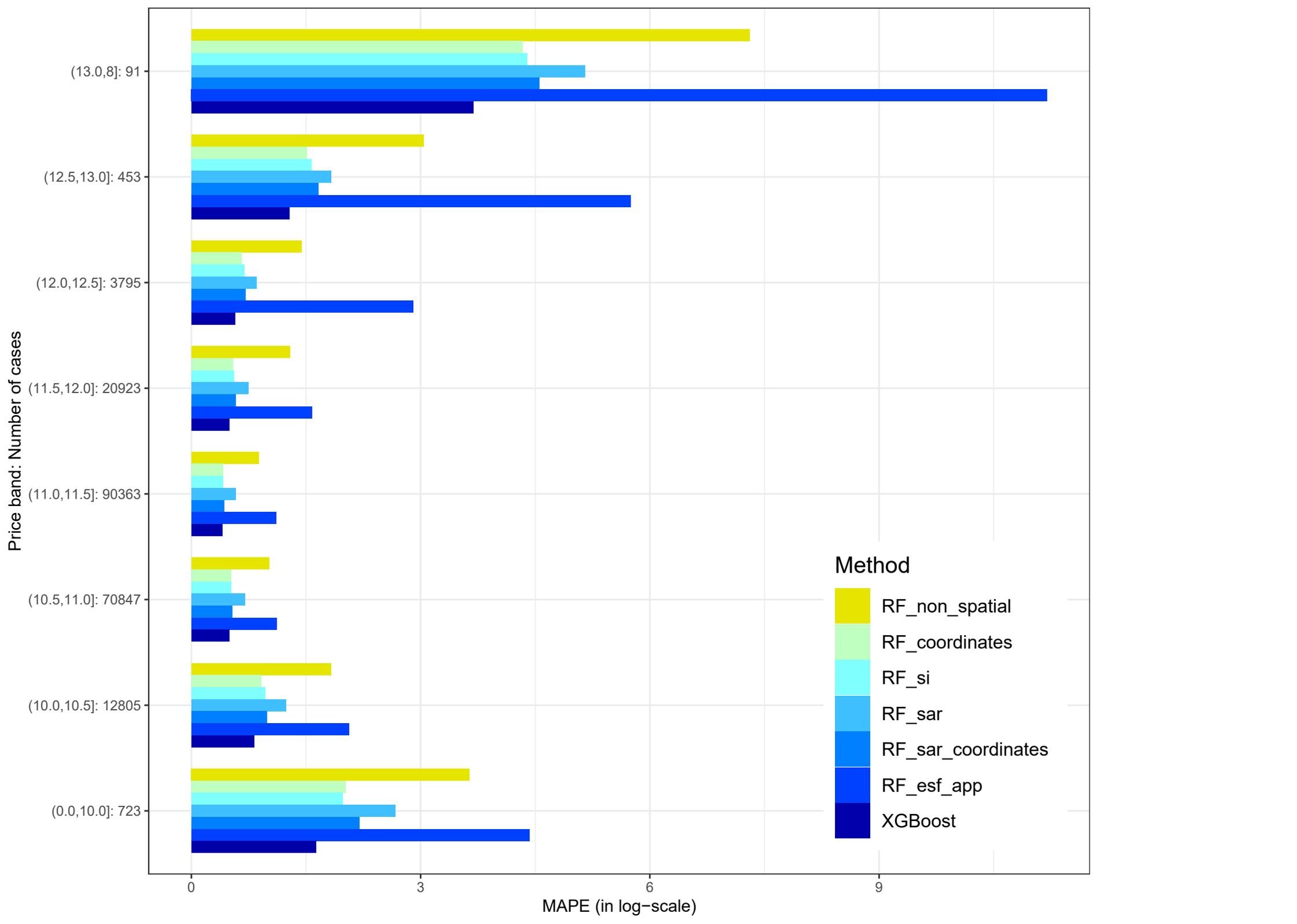